\documentclass[prc,preprintnumbers, twocolumn, superscriptaddress, amsmath, l66amssymb]{revtex4-1}

\usepackage{graphicx}% Include figure files
\usepackage{dcolumn}% Align table columns on decimal pointlip
\usepackage{bm}% bold math：
\usepackage{verbatim}
\usepackage{CJK}
\usepackage{cases}
\usepackage{amsbsy} 
\usepackage{microtype}
\usepackage{ulem}
\usepackage{tikz}
\usepackage{braket}
\usepackage{amsmath}

\def\BUAAISOE{School of Instrumentation Science and Opto-electronics Engineering, Beihang University, Beijing, 100191, China}
\def\BUAAHZ{Hangzhou Innovation Institute, Beihang University, Hangzhou, 310051, China}
\def\BUAAPhys{School of Physics, Beihang University, Beijing 100191, China}

\def\Kings{Department of Physics, King’s College London, Strand, London WC2R 2LS, UK}

\def\HIM{Helmholtz-Institut, GSI Helmholtzzentrum fur Schwerionenforschung, Mainz 55128, Germany}
\def\JGU{Johannes Gutenberg University, Mainz 55128, Germany}
\def\Berkeley{Department of Physics, University of California, Berkeley, CA 94720-7300, USA}

\usepackage{makeidx}
\begin{document}
\title{
Ultrasensitive atomic comagnetometer with enhanced nuclear spin coherence
}
\author{Kai Wei}
\affiliation{\BUAAISOE}\affiliation{\BUAAHZ}
\author{Tian Zhao}
\affiliation{\BUAAISOE}\affiliation{\BUAAHZ}
\author{Xiujie Fang}
\affiliation{\BUAAHZ}\affiliation{\BUAAPhys}
\author{Zitong Xu}
\affiliation{\BUAAISOE}\affiliation{\BUAAHZ}
\author{Chang Liu}
\affiliation{\BUAAISOE}\affiliation{\BUAAHZ}
\author{Qian Cao}
\affiliation{\BUAAISOE}\affiliation{\BUAAHZ}
\author{Arne Wickenbrock}
\affiliation{\HIM}\affiliation{\JGU}
\author{Yanhui Hu}
\email[Corresponding author: ]{yanhui.hu@kcl.ac.uk}
\affiliation{\Kings}
\author{Wei Ji}
\email[Corresponding author: ]{wei.ji.physics@gmail.com}
\affiliation{\HIM}\affiliation{\JGU}
\author{Dmitry Budker}
\affiliation{\HIM}\affiliation{\JGU}\affiliation{\Berkeley}

\date{\today}
\begin{abstract}

Achieving high energy resolution in spin systems is important for fundamental physics research and precision measurements, with alkali-noble-gas comagnetometers being among the best available sensors. We found  a new relaxation mechanism in such devices, the gradient of the Fermi-contact-interaction field that dominates the relaxation of hyperpolarized nuclear spins. We report on precise control over spin  distribution, demonstrating  a tenfold increase of nuclear spin hyperpolarization and transverse coherence time with optimal hybrid optical pumping. Operating in the self-compensation regime, our $^{21}$Ne-Rb-K comagnetometer achieves an ultrahigh inertial rotation  sensitivity of $3\times10^{-8}$\,rad/s/Hz$^{1/2}$ in the frequency range from 0.2 to 1.0 Hz,
which is equivalent to the energy resolution of $3.1\times 10^{-23}$\,eV/Hz$^{1/2}$. We propose to use this comagnetometer to search for exotic spin-dependent interactions involving proton and neutron spins. The projected sensitivity surpasses the previous experimental and  astrophysical limits by more than four orders of magnitude.

\end{abstract}
\maketitle
%\tableofcontents 

%=================================================================
%=================================================================
%=================================================================
% \section{Introduction}
%\paragraph{Introduction.--}
%=================================================================
% =================================================================
%=================================================================

Coherent control of electron and nuclear spins via light-matter interactions is an important platform for fundamental physics research\,\cite{terrano2021comagnetometer}, and an essential tool for quantum sensors\,\cite{pezze2018quantum,budker2007optical,kornack2005nuclear} and quantum information processing\,\cite{shaham2022strong}. Dense mixture of vapors of polarized alkali-metal atoms and noble gases with  hyperpolarized nuclei have found prominent use in quantum-technology devices such as atomic magnetometers and comagnetometers. Particularly, atomic comagnetometers with  hybrid spin ensembles are used to search for ``new physics'', including fifth forces \cite{lee2018improved,ji2018new}, axion-like particles \cite{bulatowicz2013laboratory,afach2021search}, permanent electric dipole moments \cite{Rosenberry2001atomic,sachdeva2019new}, and to test thee combined charge-parity-time (CPT) and Lorentz symmetries \cite{brown2010new,Smiciklas2011new}.

These applications have long been limited by systematic errors and spurious signals due to magnetic field from ambient environment or interactions between atoms \cite{sheng2014new,wu2018nuclear}. A typical approach for addressing this problem is to isolate the magnetic-field effect by using two species with different gyromagnetic ratio, for example, $^{129}$Xe and $^{131}$Xe \cite{bulatowicz2013laboratory,donley2010nuclear}, $^{3}$He and $^{129}$Xe \cite{Rosenberry2001atomic,sachdeva2019new}, $^{85}$Rb and $^{87}$Rb \cite{kimball2017constraints}, different nuclear spins in the same molecule \cite{wu2018nuclear} or different hyperfine levels of single-species atoms \cite{wang2020single}. Another approach is operating the alkali-noble gas atomic comagnetometer in the self-compensation (SC) regime\,\cite{kornack2005nuclear,kornack2002dynamics}, where noble gas nuclear spins interact with alkali electron spins by spin-exchange (SE) interactions and adiabatically cancel slowly changing magnetic fields. Another advantage of SC comagnetometer is that the alkali atoms are in the spin-exchange relaxation free (SERF) regime, one can  achieve sub-femtotesla magnetic sensitivity \cite{kominis2003subfemtotesla}.

The combination of the SC and SERF regimes enables ultrasensitive measurements of non-magnetic fields and rotations.  However, the SC mechanism for different alkali-noble-gas pairs which are used for numerous new-physics searches and inertial navigation  \cite{terrano2021comagnetometer} varies significantly and is not fully explored. The long-coherence-time $^3$He-alkali pair is used to achieve the highest sensitivity in ``fifth force'' measurements \cite{vasilakis2009limits,ji2018new}, whose SC regime breaks down at higher frequencies. In order to improve the sensitivity for non-magnetic field  or to measure exotic interactions coupling to spin-3/2 nuclei, $^{21}$Ne atoms are used instead of the $I=1/2$ $^{3}$He\,\cite{kornack2005nuclear,Smiciklas2011new}. Due to the stronger Fermi contact interaction (FCI) between $^{21}$Ne atoms and alkali atoms where the FCI field experienced by $^{21}$Ne due to Rb is two orders of magnitude higher than that for the $^{3}$He-K pair \cite{ghosh2010measurement,wei2020broadening} the SC regime is complicated by the strong FCI field from alkali atoms and the quadrupole relaxation as compared to the ``simpler'' $^{3}$He atoms.  The SC regime of heaviest stable-noble-gas $^{129}$Xe, promising for electric dipole moment measurements \cite{seltzer2008developments} and quick-start gyroscope,  is significantly influenced by the larger FCI factor $\kappa_0$ (two orders of magnitude larger than that for $^3$He) and shorter coherence time.    Moreover,  the optimal sensitivity in SERF regime is achieved when the collective spin polarization of alkali atoms is half of full polarization \cite{kominis2003subfemtotesla,dang2010ultrahigh}. This special condition results in a significant polarization gradient of alkali atoms due to strong absorption of pump light \cite{kornack2005test,brown2011new}, which further complicates the mechanism of SC suppression.

In this work, we demonstrate an ultrahigh non-magnetic field sensitivity of $3×10^{-8}$ rad/s/Hz$^{1/2}$ in the low frequency range in a SC $^{21}$Ne-Rb-K comagnetometer, which is achieved by investigating new mechanism of SC regime. To overcome the newly found relaxation mechanism of noble-gas nuclear spins that significantly shortens the coherence time and deteriorates the SC  performance,
the influences of hybrid alkali atoms (Rb-K) on spin-polarization homogeneity, hyperpolarization efficiency, and relaxation of noble-gas nuclear spins have been theoretically modelled and experimentally optimized, yielding a tenfold increase of coherence time of nuclear spins and the SC suppression ability of the hybrid comagnetometer. The energy sensitivity of this device for exotic field coupling to nuclear spins is on the order of $10^{-23}$\,eV/Hz$^{1/2}$, which is six orders of magnitude better than state-of-art comagnetometers based on Rb atoms\cite{wu2018nuclear,wang2020single}. This work will boost the experiments to search for exotic spin-dependent forces coupled to proton and neutron spins, whose sensitivity could be more than four orders of magnitude higher than previous experimental results and astrophysical limits.

\begin{figure}
\begin{center}
\includegraphics[width=8.5cm]{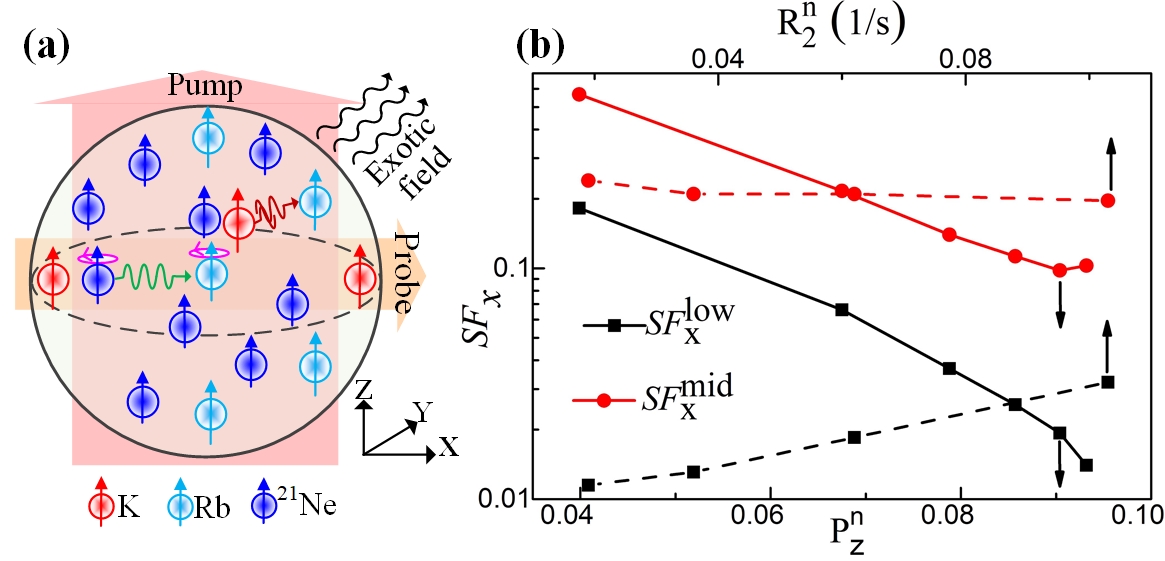}
\caption{ 
(a) The pump and probe configuration of $^{21}$Ne-Rb-K comagnetometer. Hybrid Rb-K atoms are applied to transfer the spin momentum of pumping-light photons to  $^{21}$Ne nuclear spins. The precession of $^{21}$Ne spins due to exotic fields or inertial rotation is transferred to the overlapping alkali spins, which are further read out by probe light based on optical rotation.  (b) Suppression factor $SF_x$ as a function of the noble-gas nuclear spin polarization $P^n_z$ (lower axis) and the noble-gas-spin transverse relaxation rate $R^n_2$ (upper axis). The  $SF_x^{low}$ and $SF_x^{mid}$ decrease with $P^n_z$, and the $SF_x^{low}$ increases with $R^n_2$, which are in agreement with theoretical model.   }
\label{Fig.Prin.SF} 
\end{center}
\end{figure}

\textit{Hybrid Spin Ensembles Interactions} Hybrid SC comagnetometer consists of gaseous mixture of alkali-metal  atoms and  noble-gas atoms occupying the same glass cell as illustrated in Fig.\,\ref{Fig.Prin.SF}(a). Using hybrid spin-exchange optical pumping (HSEOP), the lower-density alkali species is optically pumped and is used to polarize the higher-density alkali species via SE collisions. Due to the sufficiently large SE cross section between the hybrid alkali atom pair,  hybrid alkali spins  are strongly coupled and are in spin-temperature equilibrium with the same polarization \cite{babcock2003hybrid,lee2019new}.   Simultaneously, electron-spin polarization of the alkali atoms is transferred to noble-gas nuclear spins through SE collisions between the alkali atoms and the noble-gas atoms \cite{babcock2003hybrid}, resulting in hyperpolarization of the noble-gas nuclear spins. Under a small external magnetic field, alkali atoms  work in the  SERF regime. The spin ensembles are pumped along $\hat{z}$. The transverse polarization of higher-density alkali spins induced by the excitation signals, coupled to noble-gas spins or alkali spins, is measured by detecting optical rotation of the linearly polarized probe light propagating along $\hat{x}$.

The SE interaction between  alkali  electron spins and noble-gas nuclear spins  couples them together,  which  can be described by the FCI field seen by one spin species due to the magnetization of the other \cite{kornack2002dynamics,wei2020simultaneous}:

\begin{eqnarray}\label{eq.Effec_B}
 {\bf{{\tilde{B}}^{n/e}}}  = \frac{2}{3} \kappa _0^{\rm{e - n}}  M_0^{\rm{n/e}} {\bf{ P^{n/e}}}\,,
\end{eqnarray}
where the superscripts ``e" and ``n"  denote electron and nuclear spins, respectively;  $\kappa_0^{\rm{e-n}}$ is the  FCI enhancement factor \cite{ghosh2010measurement}; ${\bf{ P^{e}}=\bf{<S_e>}}/S_e$ is the collective polarization of alkali electron spins, ${\bf{ P^{n}}=\bf{<K_n>}}/K_n$ for noble-gas  nuclear spins; $ M_0^{\rm{e}}$ and $M_0^{\rm{n}}$  are the magnetizations of alkali electrons and noble-gas nuclei for the case of full polarization. 

%=================================================================

\textit{Self-compensation degradation mechanism.}  For different alkali-noble-gas SC comagnetometer, the working regimes are significantly different. To realise an ultrahigh sensitivity of the SC comagnetometer, it is essential to  characterize the transverse magnetic field suppression in the SC regime. We define a suppression factor $SF_x$ ($SF_y$) as the ratio of the scale factors for the response to magnetic field $B_x$  ($B_y$) and the response to a pseudo-magnetic signal (e.g. inertial rotation $\Omega _y$). The explicit expressions can be found in the Supplemental Information of Ref.\,\cite{wei2022new}:

\begin{eqnarray}\label{eq.Suppre_fact1}
 SF_x  &=& \frac{{R_2^n  + \omega /2 + {\omega ^2}\hat{\omega}_0^e /( R_2^e \hat{\omega}_0^n )}}{{\sqrt {{(\hat{\omega}_0^n )^2} + (\hat{\omega}_0^e)^2{\omega ^2}/{{ R_2^e }^2}} }}\,, \\
  SF_y  &=& \frac{{(R_2^n)^2  + \omega ^2}}{{\hat{\omega} _0^n \sqrt {(\hat{\omega}_0^n)^2 + {(\hat{\omega}_0^e )^2}{\omega ^2}/{{ R_2^e }^2}} }}\,,\label{eq.Suppre_fact2}
\end{eqnarray}
where $\omega $ is the angular frequency of the external magnetic field.  $R^n_2$ and $R^e_2$ are the transverse relaxation rates of  noble-gas nuclear spins and alkali electron spins respectively. $\hat{\omega} _0^e=\gamma_e {\tilde{B}}^e_z$ is the electron-spin Larmor precession frequency in the FCI field ${\tilde{B}}^e_z$, while $\hat{\omega} _0^n=\gamma_n {\tilde{B}}^n_z$ is the nuclear-spin Larmor precession frequency in the FCI field ${\tilde{B}}^n_z$. The precession frequencies of coupled electron and noble-gas nuclear spins are combinations of  $\hat{\omega} _0^e$ and $\hat{\omega} _0^n$ \cite{kornack2002dynamics}.
 From Eq.\,(\ref{eq.Suppre_fact1}) and (\ref{eq.Suppre_fact2}), we can find three sub-regimes considering the critical parameters $R^n_2$ and $\omega$:

\begin{equation*}
{SF_{x} \approx}\begin{cases}
    \omega/\hat{\omega}^n_0, & R^n_2 \ll \hat{\omega}_0^n \hfill (4)\\
    R^n_2/\hat{\omega}^n_0, & R^n_2\lesssim\hat{\omega}^n_0 ,\omega<R_2^n \hfill (5a) \\
     (R^n_2+\omega)/\hat{\omega}^n_0, & R^n_2\lesssim\hat{\omega}^n_0,  \omega>R_2^n \,. \hfill (5b)
     \end{cases}
\end{equation*}

The suppression factor $SF_y$ can be derived similarly, which is approximately the square of $SF_x$. $SF_x$ is about one order of magnitude worse than $SF_y$. Improving the $SF_x$  is more important for the overall performance of the device as a detector of nonmagnetic signals (rotations or exotic fields), hence we focus on $SF_x$. Previous work \cite{kornack2005nuclear,brown2011new}  described the SC regime in case (4), because  the $R_2^n$ was considered to be small as for K-$^{3}$He.
 Here, we find that for the Rb-$^{21}$Ne system and also for Cs(Rb)-$^{129}$Xe system, case (5) primarily relevant, in significant difference with the  K-$^{3}$He system. 
 
 The case (5) can be divided in to two sub-cases. In  case (5a),  the suppression factors $SF_x$ is limited by the term $R^n_2/\hat{\omega}_0^n$. This can be understood from the fact that magnetic noise $B_\bot$ is compensated by the transverse component of noble-gas nuclear magnetization  $\tilde{B}_\bot^n$  whose amplitude is determined by {$R^n_2/\hat{\omega}_0^n$}. 
  Case (5a) is $\omega$--independent, which is contrary to the case (4) for $^3$He system. In case (5b), the  $SF_x$ is limited by the term $(R^n_2+\omega)/\hat{\omega}_0^n$, which can  be interpreted as that the higher the frequency of magnetic noise $\omega$ is, the harder the  transverse component of  $\tilde{B}_\bot^n$ to follow and compensate the magnetic noise $B_\bot$, especially for $\omega$ higher than the intrinsic resonance frequency of noble-gas atoms $\hat{\omega}_0^n$.  The ways to improve the suppression ability for both two sub-cases are to increase $\hat{\omega}_0^n$, i.e the polarization of noble gas $P^n_z$, and reduce  $R^n_2$.  In the following, we define two parameters $SF_x^{low}$ and $SF_x^{mid}$, which are the values of $SF_x$ in cases (5a) and  (5b) respectively.

The SC model is applied to different operation conditions in $^{21}$Ne-Rb-K comagnetometer.   In the bottom coordinate of  Fig.\,\ref{Fig.Prin.SF} (b), the nuclear spin polarization $P^n_z$ is improved by increase the pump light intensity. $SF_x^{low}$ and $SF_x^{mid}$ decrease with  $P^n_z$, in agreement with the dependence of $SF_x$ on $\hat{\omega}^n_0 \propto P^n_z$. In the top coordinate of Fig.\,\ref{Fig.Prin.SF} (b), when increasing the cell temperature,  the $P^n_z$ and the nuclear spin transverse relaxation rate $R^n_2$ all increase. The $R^n_2$ affects the  $SF_x$ in the low-frequency range more significantly than the $P^n_z$, leading to that $SF_x^{low}$ deteriorates with $R^n_2$ regardless of the corresponding increment of $P^n_z$. 

The relaxation rate $R^n_2$ is an important parameter for the low-frequency magnetic noise suppression. To explain the observed values of $R^n_2$, we estimated the partial rates from several known relaxation mechanisms \cite{ghosh2010measurement,ghosh2009spin}, including spin-exchange and spin-destruction collisions and magnetic field gradients, and found that their sum about $1×10^{-3}$\,s$^{-1}$ is significantly smaller than the value measured.
We find that the observed relaxation rate is, in fact, dominated by the Fermi-contact-interaction field gradient $\nabla \tilde{B}^e_z$ coming from the polarization gradient $\nabla P^e_z$ of alkali spins in the SC regime, see Eq.\,(\ref{eq.Effec_B}).
The value of $\nabla \tilde{B}^e_z$ is calculated to be tens of nT/cm, much higher than the real magnetic field gradient of $\nabla B_z \approx  2$\,nT/cm. Adding this contribution to the gradient-related  relaxation  \cite{cates1988relaxation} brings the calculated value of $R^n_2$ to agreement with the measurement. 

We attempted compensate $\nabla \tilde{B}^e_z$ by using coils to generate a uniform gradient $\nabla B_z$; however, application of such a gradient always increased the relaxation rate. We believe that this is because the effective field gradient has high-order nonuniform in HSEOP, which is caused by the combination of nonlinear absorption of pump photons, density gradients of the alkali atoms and diffusion.

\textit{Improvement of SC.} In HSEOP, the polarization gradient $\nabla P^e_z$ is mainly determined by  $\xi =  n_{\textrm{Rb}}/n_{\textrm{K}}$, the ratio of alkali number densities. Therefore, it is necessary to characterize the relationship between the polarization distribution of electron and nuclear spins and the density ratio. We take into account the  diffusion of alkali and noble-gas atoms, the inhomogeneity and attenuation of pump light, cell geometry and wall relaxation to simulate the spin-polarization distribution  using  
finite-element analysis \cite{yuchen2019pump,fink2005production}. 

\begin{figure}
\includegraphics[width=8.5cm]{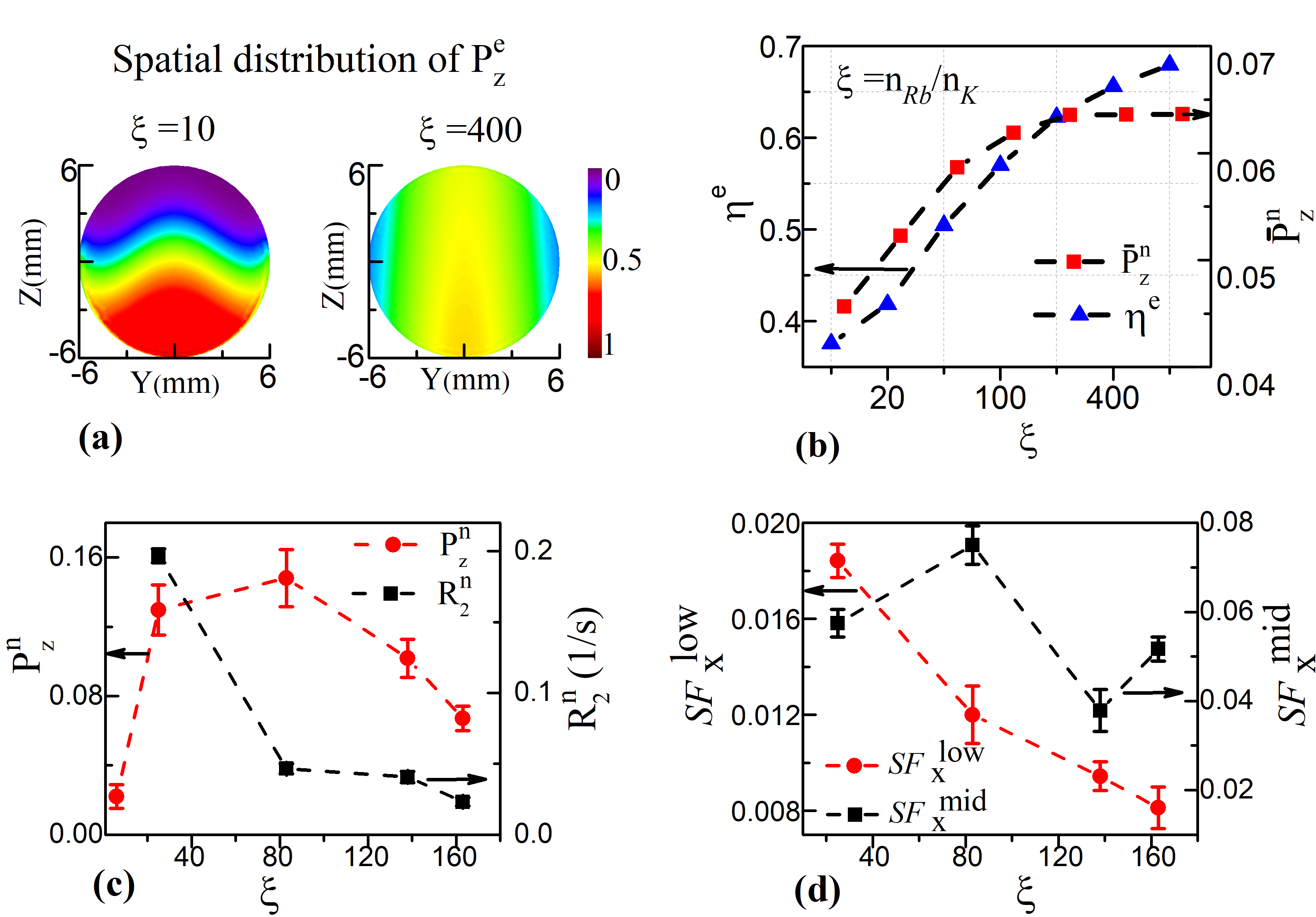}
\caption{
(a) The simulated spatial distribution of electron polarization $P^e_z$   for density ratio $\xi$ of 10 and 400 respectively in the Y-Z plane of the cell center. (b) The calculated homogeneity factor $\eta^e$  of  $P^e_z$ and the volume-averaged nuclear polarization ${\bar P}_z^n $  as functions of  $\xi$. $\eta^n\approx 1$ and $\bar{P}_z^e\approx 0.5$ are not plotted here.(c) The noble-gas spin polarization  $P^n_z$ and transverse relaxation rate $R^n_2$ as a function of the Rb-K density ratio  $\xi$ at 200 $\rm {^o}$C. By optimizing the $\xi$, the polarization $P^n_z$ and the coherence time of nuclear spins $T_2^n=1/R_2^n$ are improved by nearly one order of magnitude respectively. (d) The averaged suppression factors  $SF^{\rm{Low}}_x$ and  $SF^{\rm{Mid}}_x$  as a function of the Rb-K density ratio  $\xi$ at 200\,$\rm{^o}$C. 
}
\centering
\label{Fig.Simu.diff.Dr} 
\end{figure} 

In the simulation, $P^e_z$ at the  cell center is normalized to 0.5 at 190$^{\rm{o}}$C, where the sensitivity  of the SERF magnetometer is optimal. The pump light beam has a Gaussian profile with 18\,mm beam diameter to cover the 12\,mm diameter spherical vapor cell. The cell is filled with 2280\,torr (3\,amg) $^{21}$Ne and 50\,torr N$_2$. Other parameters of the $^{21}$Ne-Rb-K spin ensembles are the same as in Refs.\,\cite{ghosh2009spin,lee2019new}. As shown in Fig.\,\ref{Fig.Simu.diff.Dr}(a), for a small $\xi$, $P^e_z$ decreases significantly along $\hat{z}$ (along the pump-light propagation direction) , while for a larger $\xi$, $P^e_z$ becomes more uniform. In comparison, $P^n_z$ is always spatially homogeneous, due to the fact that the diffusion rate of noble gas is faster than its relaxation rate. We use $\eta^e=\bar{P}_z^e/P_{\rm{z,Max}}^e$, the ratio of the volume-averaged value to the maximum value to characterize the homogeneity of the polarization. Figure\,\ref{Fig.Simu.diff.Dr}(b) shows the dependence of $\eta^e$ and the volume-averaged nuclear spin polarization ${\bar P}_z^n $ on the density ratio $\xi$. The polarization  ${\bar P}_z^n $ saturates at $\xi\approx100$ while homogeneity $\eta^e$ continues to increases.

Five $^{21}$Ne-Rb-K cells with different $\xi$ were tested, with the values of $\xi=$ 6, 25, 83, 138 and 163, respectively.  Apart from $\xi$, other parameters were kept nearly the same, i.e. $^{21}$Ne density about 2.67\,$  \sim$\,3.24\,amg and N$_{2}$ pressure about 35\,$\sim$\,53\,torr. 
As shown in Fig.\,\ref{Fig.Simu.diff.Dr}(c), the $P^n_z$ for each cell increases with $\xi$ but reaches a maximum at approximately $\xi=83$ and then decreases, which is different from the simulation result in Fig.\,\ref{Fig.Simu.diff.Dr}(b). The difference is due to the fact that $P^e_z$ in the simulations of ${\bar P}_z^n $ is set to 0.5 for different $\xi$, while in the experiment, the available pump-light intensity is insufficient for the cell with larger $\xi$ to achieve high $P^e_z$, leading to a smaller $P^n_z$. With the increase of the pump-light intensity, $P^n_z$ can increase for larger $\xi$. However, this increase breaks down for large $\xi$, because the SE between K  and Rb atoms is too slow to transfer the polarization of photon spin efficiently. Because  $P^n_z$ is small for the cell with $\xi=6$, we focus on the cells with $\xi$  ranging from 25 to 163. Figure\,\ref{Fig.Simu.diff.Dr}(c) shows the relationship between  $R^n_2$ and  $\xi$ for a range of pump-light intensities. $R^n_2$ decreases with $\xi$  due to the reduction of FCI gradient with  $\xi$, which agrees with the theoretical expectation in Fig.\,\ref{Fig.Simu.diff.Dr}(b). 

The suppression factors $SF_x$ for these four cells are shown in Fig.\,\ref{Fig.Simu.diff.Dr} (d). $SF^{\rm{Low}}_x$ is mostly dominated by $R_2^n$, while the behaviour of $SF^{\rm{Mid}}_x$ appears to be more complicated as it depends on  factors including $R_2^n$, $\omega$, $P_z^n$ and the pump-light intensity.  Although $\xi=83$ achieves the highest  $P^n_z$, its suppression ability is not the optimal, indicating that improving the noble-gas polarization alone is insufficient to optimize the suppression of magnetic noise.  The reduction of relaxation rate is relatively more important. The optimal suppression ability occurs when $\xi=163$, which is one order of magnitude higher than that of $\xi=25$. The low-frequency suppression factor for the $\xi=163$ cell is smaller than 0.01, which means the magnetic noise in the magnetic shield (on the order of fT) can be suppressed by two orders of magnitude.

\begin{figure}
\begin{center}
\includegraphics[width=8cm]{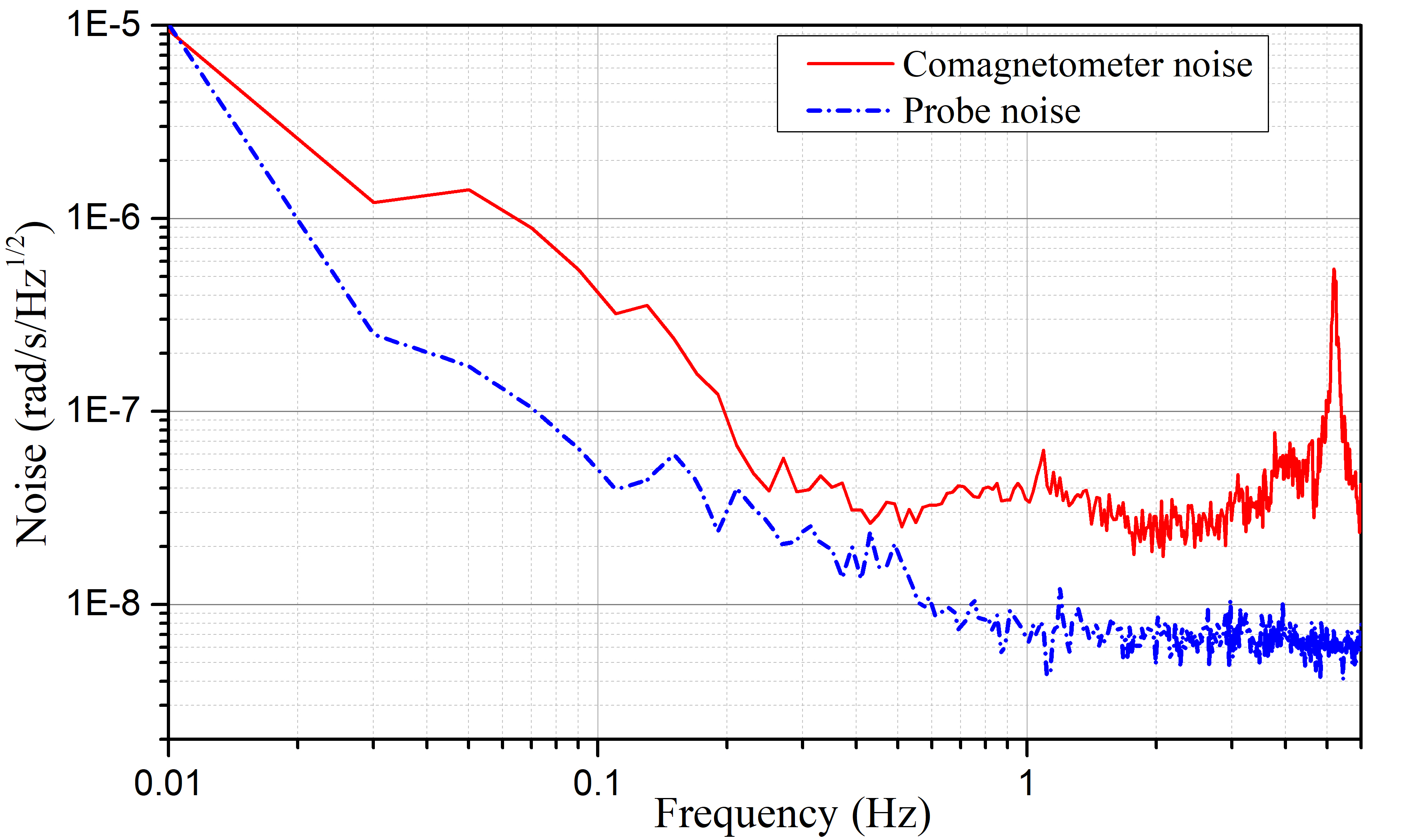}
\caption{
The noise spectrum of the SC comagnetometer. In the frequency range from 0.2 to 1.0\,Hz, the averaged noise is $3×10^{-8}$\,rad/s/Hz$^{1/2}$. Below 0.2\,Hz, the noise spectrum  is dominated by various 1/f noises. The peak in the noise spectrum is due to the vibration which is confirmed with precision seismometer. The probe noise,  which is measured with unpolarized spin ensembles by blocking the pump light,  is smaller than the comagnetometer noise and approaches the photon shot noise limit.
}
\label{Fig.Sens} 
\end{center}
\end{figure}

The amplitude spectral density of comagnetometer signal is shown in Fig.\,\ref{Fig.Sens}. The calibration of the SC comagnetometer sensitivity is done by measuring the rotation of the Earth \cite{wei2022new}. The peak around 5.0\,Hz  is related to the vibration resonance of the vibration-isolation platform, which is  confirmed with a seismometer installed on the platform. Below 0.2\,Hz, the comagnetometer noise is dominated by the $1/f$ noise caused by various drifts such as those of temperature and pump-light alignment. The measured noise spectrum reflects the ability of the device to suppress magnetic noise. The noise from the inner ferrite shield is calculated based on finite element analysis using the measured complex permeability  $\mu '/ \mu_0  = 6308(14)$, $\mu ''/ \mu_0  = 45(3)$  and geometric parameters \cite{ma2021parameter}. The magnetic noise is calculated to be about  $2.5 f^{-1/2}$\,fT ($f$ in unit of Hz) and converted to the corresponding noise with unit of rad/s/Hz$^{1/2}$ by multiplying the gyromagnetic ratio of $^{21}$Ne \cite{kornack2005nuclear}. Except for the frequency range dominated by the $1/f$ noise, the noise from the ferrite shield exceeds the comagnetometer noise in the low-frequency range, indicating that the magnetic noise is effectively suppressed by the SC effect. The probe-light noise is measured to be smaller than the comagnetometer noise. Above 1.0\,Hz, the polarimetry noise of probe light based on optical rotation is lower than $2×10^{-8}\, \rm{rad/Hz^{1/2}}$, approaching the limit of photon shot noise \cite{ledbetter2008spin}. 

The averaged noise is $3×10^{-8}$\,rad/s/Hz$^{1/2}$  in the frequency range from 0.2 to 1.0\,Hz, corresponding to an effective pseudomagnetic field sensitivity of $\delta b^{n}=1.5$\,fT/Hz$^{1/2}$. For the exotic-field Zeeman-like pseudomagnetic coupling to $^{21}$Ne nuclear spin, the energy is $E=\mathbf{\mu}_{Ne} \cdot \mathbf{b}^n$, yielding an energy resolution of $\delta E_{Ne}=3.1\times 10^{-23}$ eV/Hz$^{1/2}$. 

The  energy sensitivity of the exotic field coupling to neutron and proton spins are determined by $\delta E_{n/p}=\delta E_{Ne}/\eta_{n/p}$, where $\eta_n=0.58$  and $\eta_p=0.04$ are the neutron and proton fraction of spin polarization in $^{21}$Ne atoms \cite{almasi2020new} respectively. Therefore, the energy sensitivity of our setup are $\delta E_n=5.4\times 10^{-23}$ eV/Hz$^{1/2}$ and $\delta E_p=7.8 \times 10^{-22}$ eV/Hz$^{1/2}$ respectively.

Based on the demonstrated performance of the device, we propose an experiment to search for exotic interaction between the comagnetometer spins and the Earth gravitational field, using the geometry similar to that of Ref.\,\cite{venema1992search,kimball2017constraints}. The spin-dependent force could be mediated, for example, by an ultralight spin-0 boson, such as axion or axionlike particle (ALP) \cite{moody1984new,dobrescu2006spin,fadeev2019revisiting}: 
\begin{equation*}
    \begin{aligned}
V=&\frac{ g_S g_P \hbar^2}{8 \pi m}(\hat{\boldsymbol\sigma} \cdot \hat{\boldsymbol r})
\left(\frac{1}{r\lambda}+\frac{1}{r^2}\right)e^{-r/\lambda} ,\hfill (6)  \\ \end{aligned}
\end{equation*}
where $g_s$ and $g_p$ are scalar and pseudoscalar coupling constants; $\hbar$ is the reduced Planck constant; $\hat{\sigma}$ is the Pauli spin-matrix vector of one fermion and $m$ is its mass; $\lambda$ is the force range, which is inversely proportional to the mass of the force-mediating boson; $r$ is the relative distance between two fermions and $\hat{r}$ is the unit vector directed from the one fermion to the other. If it exists, this exotic force violates parity (P) and time-reversal invariance (T).

\begin{figure}
\begin{center}
\includegraphics[width=8cm]{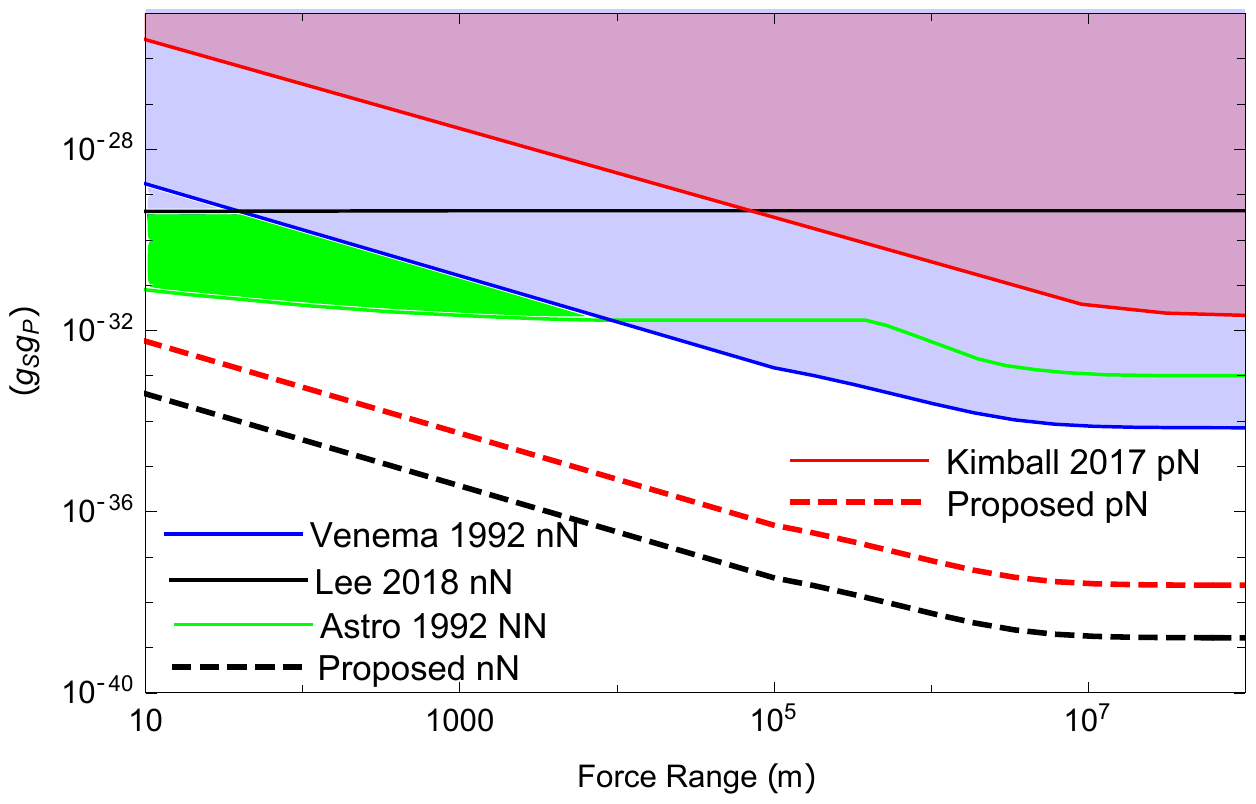}
\caption{ Existing limits and projected sensitivity on the coupling-constant product $g_S g_P$ of the spin-dependent force between the neutron (proton) spin and unpolarized nucleon. The black dashed and red dashed lines are the sensitivity in this proposal using the $^{21}$Ne neutron spin and proton spin, respectively. The blue solid line ``Venema 1992 nN" \cite{venema1992search} and black solid line ``Lee 2018 nN" \cite{lee2018improved} represent the direct experimental limits on the coupling to neutron spins, while the red solid line ``Kimball 2017 pN" \cite{kimball2017constraints} represents the direct experimental limits on the coupling to proton spins. The green solid line ``Astro 1992 NN"  \cite{raffelt2012limits} represents the astrophysical limits on the coupling between nucleons (not distinguishing between protons and neutrons).
}
\label{Fig.limits} 
\end{center}
\end{figure}

The state-of-art experiments for the proton spin-gravity coupling include the $^{85}$Rb and $^{87}$Rb comagnetometer\,\cite{kimball2017constraints}  or the $^{87}$Rb comagnetometer using  two hyperfine levels of $^{87}$Rb  atoms\,\cite{wang2020single}. Both these experiment realized energy resolutions on the order of $10^{-18}$\,eV for an integration time of more than one hundred hours. The best experimental result for the neutron-spin coupling to Earth gravity was obtained with a $^{199}$Hg-$^{201}$Hg comagnetometer and realized an energy resolution on the order of $10^{-21}$ eV\cite{venema1992search}. 
We estimate the sensitivity of our experiment using Earth as a source and integrating for about 100 hours using a similar approach to that work of Ref.\,\cite{venema1992search,kimball2017constraints,Smiciklas2011new}. The estimated exotic magnetic field sensitivity is $\delta B 	\lesssim 0.01\, \rm{fT}$, and energy sensitivity as $\delta E_n\lesssim 4\times 10^{-25}\,\rm{eV}$ and $\delta E_p\lesssim5\times 10^{-24}$\,eV respectively. The estimated sensitivity to the coupling constants and a comparison with the previous work is shown in Fig.\,\ref{Fig.limits}.  The sensitivity of this proposal can surpass the direct experimental limits and the astrophysical limits on the exotic interactions coupling to proton and neutron spins by more than four orders of magnitude.

Taking advantage of the ultrahigh sensitivity, this comagnetometer can also be used to explore new spin-dependent physics, including directly searching for axion and  axion-like particles (ALPs) \cite{afach2021search}, and local Lorentz invariance (related to the CPT symmetry) \cite{brown2010new,Smiciklas2011new}. 

%=================================================================
% \section{Discussion and conclusions}
%=================================================================

 \textit{Conclusion} We establish an analytical model to  describe the SC effect and find that the relaxation and polarization of noble-gas nuclear spins are the key factors determining the SC performance.  The relaxation of noble-gas nuclear spins is found to be dominated by a new relaxation mechanism, i.e. the Fermi-contact-interaction field gradient resulting from nonuniform alkali spin polarization.  The degree of spin polarization and its relaxation time for the noble-gas atoms are both improved by one order of magnitude  over the earlier work \cite{pang2022highly} by optimizing the density ratio between the two alkali species.  An average sensitivity of 3×10$^{-8}$\,rad/s/Hz$^{1/2}$ in the frequency range from 0.2\,Hz to 2.0\,Hz has been achieved. This sensitivity represents six orders of magnitude better energy resolution compared to comagnetometers using Rb isotopes that were used to search for exotic gravity coupling to proton spins \cite{kimball2017constraints}. 
 The improvement of nuclear spin coherence time and polarization  is also beneficial for NMR gyroscopes, noble-gas-spins-based quantum memory,  coherent bidirectional coupling between light and noble-gas spins \cite{shaham2022strong}, as well as for neutron spin filters\cite{zhang2022situ}. 

%=================================================================
%=================================================================

%==================================================
%==================================================

%===============

\begin{acknowledgments}

This work is supported
by the National Natural Science Foundation of China
(NSFC) (Grant Nos. 62203030 and 61925301 for Distinguished Young
Scholars), the China postdoctoral Science
Foundation (Grant No. 2021M700345), the DFG Project
ID 390831469: EXC 2118 (PRISMA+ Cluster of Excel-
lence), by the German Federal Ministry of Education
and Research (BMBF) within the Quantumtechnologien
program (Grant No. 13N15064), and by the QuantERA
project LEMAQUME (DFG Project Number 500314265).

\end{acknowledgments}

%\end{CJK*}
\bibliographystyle{apsrev4}
\bibliography{Ultrasens}

\end{document}